\documentclass[aps,pra,amsmath,superscriptaddress,amssymb,twocolumn,showpacs]{revtex4-1} 
\pdfoutput=1

\usepackage{graphicx}
\usepackage{dcolumn}
\usepackage{bm}
\usepackage{bbold}
\usepackage{nonfloat}
\usepackage{color}

\newcommand{\Dollar}{\$}

\definecolor{pink}{rgb}{1,0.16,0.64}
\definecolor{brown}{rgb}{0.65,0.16,0.16}
\definecolor{purple}{rgb}{0.5,0,0.5}
\definecolor{grey}{rgb}{0.3,0.3,0.3}
\definecolor{green2}{rgb}{0,0.5,0}

\begin{document}

\preprint{PRE/004} 

\title{Quantum error correction and detection:\\quantitative analysis of a coherent-state amplitude damping code}

\author{Ricardo Wickert}
 \email{ricardo.wickert@mpl.mpg.de}
\affiliation{Optical Quantum Information Theory Group, Max Planck Institute for the Science of Light, G\"unther-Scharowsky-Str. 1/Bau 26, 91058 Erlangen, Germany}
\affiliation{Institute of Theoretical Physics I, Universit\"at Erlangen-N\"urnberg, Staudttr. 7/B2, 91058 Erlangen, Germany}

\author{Peter van Loock$^{1,2,}$}%
\affiliation{Universit\"at Mainz, Staundingerweg 7, 55128 Mainz, Germany}

\date{\today}


\begin{abstract}
We re-examine a non-Gaussian quantum error correction code designed to protect optical coherent-state qubits against errors due to an amplitude damping channel. We improve on a previous result [Phys. Rev. \textbf{A} 81, 062344 (2010)] by providing a tighter upper bound on the performance attained when considering realistic assumptions which constrain the operation of the gates employed in the scheme. The quantitative characterization is performed through measures of fidelity and concurrence, the latter obtained by employing the code as an entanglement distillation protocol. We find that, when running the code in fully-deterministic error correction mode, direct transmission can only be beaten for certain combinations of channel and input state parameters, whereas in error detection mode, the usage of higher repetition encodings remains beneficial throughout.
\end{abstract}

\maketitle

\section{Introduction} 

The ability to safeguard entanglement and coherence in fragile quantum states is a fundamental element to any future quantum computing architecture or communication network. Quantum Error Correction (QEC) \cite{Shor,*GottesmanPhD} enables such task by encoding the quantum information carrier in a larger, more robust Hilbert space. Alternatively, Entanglement Purification (EP) \cite{duerbriegel} protocols rely on distilling entanglement from a number of identically-prepared copies of the state; both approaches have been shown to be interchangeable  \cite{Bennett,CerfNoGo,Aschauer}.

In optical implementations of Quantum Information Processing (QIP) protocols, photon loss, or amplitude damping, is the predominant source of decoherence. It is modeled by a beam splitter interaction of the signal with a vacuum mode, with an appropriately chosen transmissivity parameter. This corresponds to a Gaussian error, and as such, Gaussian operations alone - i.e., those implementable through Gaussian ancillae, beam splitters, phase-shifters and homodyne measurements - do not suffice to protect the signal state \cite{CerfNoGo,Giedke,*Eisert,*Fiurasek}, and more elaborate, non-Gaussian operations must be employed.

Given that, typically, such non-Gaussian elements can only be implemented non-deterministically, a further distinction becomes necessary:  We denote as operating in \textit{Error Correction mode} those codes which unconditionally produce an output state at every execution; alternatively, those codes carried out in a post-selected fashion - such as the heralded-noiseless-amplifier scheme proposed by Ralph in \cite{RalphHLA} - are said to be run under in \textit{Error Detection mode}. Whereas in the first scenario all communication flows in a single direction, enabling true, deterministic error correction, the latter requires two-way communication, \textit{i.e.}, the receiving party must acknowledge a successful run, or indicate that the input state must be transmitted anew.

In a previous work \cite{OurPaper}, the QEC scheme of ref. \cite{GVR} was revisited. This scheme protects coherent-state superpositions (CSS, also known as ``cat states" \cite{RalphLund}) through the use of non-Gaussian encoding operations. A quantitative analysis was provided by examining the code from an entanglement distillation perspective, employing Wootters' concurrence \cite{Wooters}.

Here, we offer a two-fold expansion of the previous result. First, by operating the error correction protocol with single-mode input states, we calculate different fidelity-based measures. Together with the entanglement-based analysis employed previously, this allows us to quantitatively evaluate the performance of codes both as an error correction scheme and as an entanglement purification protocol.  Then, we address the fundamental non-unitarity of the involved Hadamard gates, lifting the previous notion that assumed a satisfactory realization of this operation even for fairly small superposition sizes. The non-unitarity of the gate is addressed either via a post-selection step, which turns the error correction scheme into a probabilistic protocol, or by means of an equivalent unitary transformation which, while enabling the code to be run deterministically, introduces an undesired component in the target state. In particular when working in this fully-deterministic error correction mode, the results indicate a remarkable, and perhaps counter-intuitive, trade-off feature of these protocols.

The remainder of this paper is structured as follows: in Section II we briefly review the coherent-state alphabet which will serve as input to our encoding. Section III introduces the reader to coherent-state phase-flip code; the operation of the Hadamard gates in this protocol is discussed in Section IV. Section V discusses the connection between non-unitary, probabilistic gates and different error correction regimes. Performance measures used to quantify the performance of the scheme are presented in Section VI; with the measures being subsequently employed to obtain the results in Section VII. Concluding remarks are brought forth in Section VIII.

\section{Coherent-State qubits} 

A coherent-state alphabet identifies the logical states as  $|0\rangle_L = |-\alpha\rangle$ and $|1\rangle_L = |\alpha\rangle$. Here, $|\alpha\rangle$ is a coherent-state defined as an eigenvalue of the annihilation operator ($\hat{a} |\alpha\rangle = \alpha |\alpha\rangle$, with $\alpha \in \mathbb{C}$) and expressed in the Fock (number) basis as
\begin{align}
\label{coherentfock}
|\alpha\rangle = e^{-|\alpha|^2} \sum_{n=0}^{\infty} \frac{\alpha^n}{\sqrt{n!}}|n\rangle \quad .
\end{align}
An arbitrary qubit is therefore represented as
\begin{align}
\label{coherentqubit}
|Q\rangle = \frac{1}{\sqrt{N(\alpha)}} ( \sqrt{w} |-\alpha\rangle + e^{i \theta} \sqrt{1-w} |\alpha\rangle ) \quad ,
\end{align}
where $0 \geq w \geq 1$, $0 \geq \theta \geq \pi$, and $N(\alpha)$ is a normalization constant, $N(\alpha) = 1 + 2 \cos \theta \sqrt{w(1-w)} e^{-2|\alpha|^2}$. We note that for large values of $|\alpha|$, $|-\alpha\rangle$ and $|\alpha\rangle$ are approximately orthogonal and $N(\alpha) \approx 1$. However, currently only superpositions of limited $|\alpha|$ sizes can be attained (``Schr\"odinger Kittens" \cite{GrangierScience}). Thus, a significant amount of non-orthogonality must be considered.

The photon loss channel $\$$ is modeled by letting the signal interact with a vacuum mode $|0\rangle_l$ in a beam splitter of transmissivity $\eta$, resulting in
\begin{align}
\label{bssup}
|Q\rangle_{\$} = \frac{1}{\sqrt{N(\alpha)}} &( \sqrt{w} |-\alpha\sqrt{\eta}\rangle|-\alpha\sqrt{1-\eta}\rangle_l \nonumber \\
& + \; e^{i \theta} \sqrt{1-w} |\alpha\sqrt{\eta}\rangle|\alpha\sqrt{1-\eta}\rangle_l ) \quad .
\end{align}
From (\ref{bssup}) one obtains the final state after transmission by integrating over the loss mode (denoted here by $|\beta\rangle_l$):
\begin{align}
\label{trace}
\rho = \frac{1}{\pi}\int {} d^2\beta \,\: _l\langle\beta|Q\rangle_{\$} {} _{\$}\langle Q| \beta \rangle_l \quad .
\end{align}
Although for a single coherent state this integration is trivial and amounts to an amplitude contraction, for a superposition the resulting state after tracing out the loss mode is a mixture. One obtains
\begin{align}
\label{tracemixed2}
\rho_Q = (1-p_e)|Q_{\alpha\sqrt{\eta}}\rangle \langle Q_{\alpha\sqrt{\eta}}| + p_e Z|Q_{\alpha\sqrt{\eta}}\rangle \langle Q_{\alpha\sqrt{\eta}}|Z \; ,
\end{align}
where $p_e = \frac{1}{2}[1-e^{-2(1-\eta)|\alpha|^2}]$ is the probability that the Pauli Z operator, $Z(a|0\rangle_L + b|1\rangle_L) = a|0\rangle_L - b|1\rangle_L$, was applied. It approaches $\frac{1}{2}$ in the limit of large $\alpha$ for any $\eta < 1$. The two-fold effect of photon loss becomes explicitly manifest in this expression: first, the amplitude of the states is unconditionally reduced from $\alpha$ to $\alpha \sqrt{\eta}$; second, with probability $p_e$, the qubit suffers a phase flip.

\section{GVR Error Correction Scheme} 

The Glancy-Vasconcelos-Ralph (GVR) code \cite{GVR} is a continuous-variable adaptation of a qubit phase-flip code, designed to protect quantum information carriers encoded in a binary coherent-state alphabet. Protection against the undesired effects of an amplitude-damping induced phase-flip channel is granted to an optical qubit by transmitting it through a sequence of three beam splitters, followed by Hadamard gates - a highly non-Gaussian operation which, ideally, implements $|0\rangle_L \rightarrow |0\rangle_L + |1\rangle_L$, $|1\rangle_L \rightarrow |0\rangle_L - |1\rangle_L$, up to a normalization constant. Deviations from this ideal behaviour due to the non-orthogonal nature of the logical qubits will be addressed in the following section. The (unnormalized) encoded state which results is
\begin{align}
\label{encoded}
\sqrt{w} ( |-\alpha\rangle + |\alpha\rangle )^{\otimes 3} + e^{i \theta} \sqrt{1-2} ( |-\alpha\rangle - |\alpha\rangle )^{\otimes 3}
\end{align}

After transmission through the lossy channels, another set of Hadamard gates is in order: the transformation is applied to each of the modes, which are then recombined through an inverted sequence of beam splitters. The two ancilla modes are measured to provide syndrome information, from which the appropriate correcting operation can be applied to return the signal to its ``unflipped" state. 

The three-way redundant encoding achieved by the procedure outlined above can correct up to one error (\textit{i.e.}, a single phase flip in any of the three channels); error-free transmission is achieved with a probability given by
\begin{equation}
\label{Psuccess}
p_{success,3} = 1 - 3 p_e^2 + 2 p_e^3 \quad .
\end{equation}
In other words, similar to the original qubit code, the effective error probability is reduced from $p_e$ to $p_e^2$. This success rate can be increased by encoding the input state in a larger number of modes; to provide insight on the value of such higher-order codes, schemes based on 5, 11, and 51 repetitions will be also considered. A code based on $N$ repetitions achieves 
\begin{equation}
\label{pgeneral}
p_{success,N} = \sum_{k=0}^{\frac{N-1}{2}} \binom{N}{N-k} (1-p_e)^{N-k} p_e^k  \quad ,
\end{equation}
with an effective error probability reduced to $p_e^{\frac{N+1}{2}}$.

\section{Coherent-state Hadamard gate}

As pointed out earlier, the Hadamard gate is defined as a unitary operation which, up to normalization, effects the transformation
\begin{align}
\label{idealhadamard} 
|0\rangle_L &\stackrel{\hat{H}}{\rightarrow} |0\rangle_L + |1\rangle_L \quad , \nonumber \\
|1\rangle_L &\stackrel{\hat{H}}{\rightarrow} |0\rangle_L - |1\rangle_L \quad .
\end{align}
However, when working with a coherent-state basis, one must pay special attention to the non-orthogonal nature of the logical alphabet. If one blindly substitutes $|0\rangle_L = |-\alpha\rangle$ and $|1\rangle_L = |\alpha\rangle$ in the expression above, at the same time assuming the unitarity of the Hadamard gate, i.e., $\hat{H}^{\dagger} \hat{H}=\mathbb{1}$, one is led to a contradiction. Namely, 
\begin{align}
_{L}\langle 0 | 1 \rangle_L &= \langle -\alpha | \alpha \rangle = e^{-2|\alpha|^2} \quad ,\\
\mbox{whereas } \quad \quad \quad \quad \quad \quad & \nonumber \\
_{L} \langle 0 | \hat{H}^{\dagger} \hat{H} | 1 \rangle _L & \propto \big({}_{L}\langle 0| + {}_{L}\langle 1| \big) \big( |0\rangle_{L} - |1\rangle_{L} \big)  \nonumber \\
&\propto \langle Q_{+} | Q_{-} \rangle = 0 \quad ,
\end{align}
where $|Q_{\pm}\rangle = \frac{|-\alpha\rangle \pm |\alpha\rangle}{\sqrt{N_{\pm}(\alpha)}}$, and $N_{\pm}(\alpha) = 2 \pm 2e^{-2|\alpha|^2}$.

Clearly, a unitary transformation must preserve overlaps. This is achieved when one defines the action of the Hadamard gate in terms of an orthogonal basis $\{ |u\rangle,|v\rangle \}$. The codewords are rewritten as
\begin{align}
\label{orthobasis}
|-\alpha\rangle = \mu_\alpha |u_{\alpha}\rangle& - \nu_\alpha |v_{\alpha}\rangle \quad \mbox{ and} \nonumber \\
|\alpha\rangle = \mu_\alpha |u_{\alpha}\rangle& + \nu_\alpha |v_{\alpha}\rangle \; ,
\end{align}
with $\mu_\alpha^2 = \frac{1 + e^{-2|\alpha|^2}}{2}$ and $\nu_\alpha^2 = \frac{1 - e^{-2|\alpha|^2}}{2}$ (for convenience of notation, the subscript $\alpha$ will be dropped, and a global phase is introduced), and the Hadamard acts as
\begin{align}
\label{unithadamard}
|u\rangle \quad & \stackrel{\hat{H}}{\rightarrow} \quad \quad \frac{|u\rangle - |v\rangle}{\sqrt{2}} \quad , \nonumber \\
|v\rangle \quad & \stackrel{\hat{H}}{\rightarrow} \quad - \frac{|u\rangle + |v\rangle}{\sqrt{2}} \quad . 
\end{align}
For large values of $\alpha$, we recover Eq. (\ref{idealhadamard}) exactly; however, for those smaller values where taking this limit is not valid, using Eqs. (\ref{orthobasis}) and (\ref{unithadamard}), we obtain 
\begin{align}
\label{hadamardlong}
\hat{H}|\pm \alpha \rangle =& \mu \left( \frac{|u\rangle - |v\rangle}{\sqrt{2}} \right) \pm \nu\left( - \frac{|u\rangle + |v\rangle}{\sqrt{2}} \right) \nonumber \\
=&  \frac{1}{\sqrt{2}} \left[ \left(\mu \mp \nu\right) |u\rangle - \left(\mu \pm \nu\right) |v\rangle \right] \nonumber \\
=&  \frac{1}{\sqrt{2}} \left[ \left(\mu \mp \nu\right) |Q_{+}\rangle + \left(\mu \pm \nu\right) |Q_{-}\rangle \right]  \quad . 
\end{align}
One can then identify $\frac{\mu + \nu}{\sqrt{2}}$ as the ``successful" fraction of the resulting state, whereas $\frac{\mu - \nu}{\sqrt{2}}$ corresponds to the erroneous component which results from enforcing the unitary behavior. By following this gate with a QND measurement procedure in the $\{|Q_{\pm}\rangle\}$ basis, one effectively obtains a ``successful" and an ``erroneous" probability for the Hadamard operation \footnote{The QND measurement is useful for two reasons: first, it forces a flip/no flip behavior, which enables one to combine the Hadamard failure probability with the phase flip probability $p_e$ induced by amplitude damping channel, resulting in a global success probability for the code based on the triplet ($\alpha$,$\eta$,$N$). Second, by removing the off-diagonal elements in the resulting density matrix, the entanglement can be calculated by more effective means, see section VI. C.}.

\section{Probabilistic gates and error correction regimes}
Disregarding the result of the QND measurement outlined above, one can operate the error-correcting scheme in an unconditional and deterministic fashion. Alternatively, one could post-select on the favorable measurement outcomes, implementing a heralded Hadamard-like operation which only effected the desired transformation intended by the ``original" gate. This gives rise to two different scenarios: one with fully post-selected gates, and one with post-selected encoding, but deterministic decoding: the \textit{encoding} operations can be thought of an off-line resource, where the gate is only teleported into the input state \cite{GottesmanChuang} after a successful heralding has taken place. However, the same concept cannot be implemented in the \textit{decoding} part of the protocol, lest the quantum information be lost, requiring the receiver to signalize the sender that a new attempt is necessary. This incurs the need for two-way communication in the error correction toolbox, and ultimately results in a probabilistic operation of the protocol, with its rate upper bounded by the ``successful" component identified in the preceding section.

\section{Performance of the Codes}
A naive interpretation based on Eq. (\ref{pgeneral}) could lead one to conclude that higher-order codes always yield better results. However, this conclusion is sometimes erroneous: a close inspection of the results brought forth in section IV shows that the Hadamard operation has a higher probability of failure when operating on an input alphabet based on ``smaller" coherent states. As the number of repetitions increases, the states reaching the Hadamard gates decreases following $\alpha / \sqrt{N}$ (resp. $\alpha \sqrt{\eta}/ \sqrt{N}$ after the amplitude damping channel), thus adversely affecting a code's performance. Nevertheless, independently of this consideration, there are numerous ways of defining ``better", which can lead to a different ordering between codes. It is then necessary to define quantitative measures to establish proper comparisons.

\subsection{(Worst-case) Fidelity}
The prototypical benchmarking of quantum error-correcting codes is given by the fidelity between input and output states \cite{Uhlmann,*Jozsa}, \textit{i.e.},
\begin{align}
\label{fidworstcase}
F = \min_{Q} \sqrt{\langle Q | \rho_Q | Q\rangle} \quad .
\end{align}
where $\rho_Q = \Dollar \left( |Q\rangle\langle Q| \right)$ corresponds to the state after action of the channel $\Dollar$ (potentially including encoding and decoding operations). The minimization is performed over the whole range of input states. In the code presently being considered, a different worst-performing state is found for different values of the damping constant and/or coherent-state superposition sizes. This behaviour makes for laborious comparisons, as one has to keep track, at each pair ($\eta$,$\alpha$), which parameters $w$ and $\theta$ in Eq. (\ref{coherentqubit}) correspond to the worst-performing input state (see Fig. 1). For this reason, this measure will not be employed in our subsequent analysis, with preference given to the two following quantities.

\quad \\

\noindent\begin{minipage}{1\linewidth}
\centering
\includegraphics[scale=0.51]{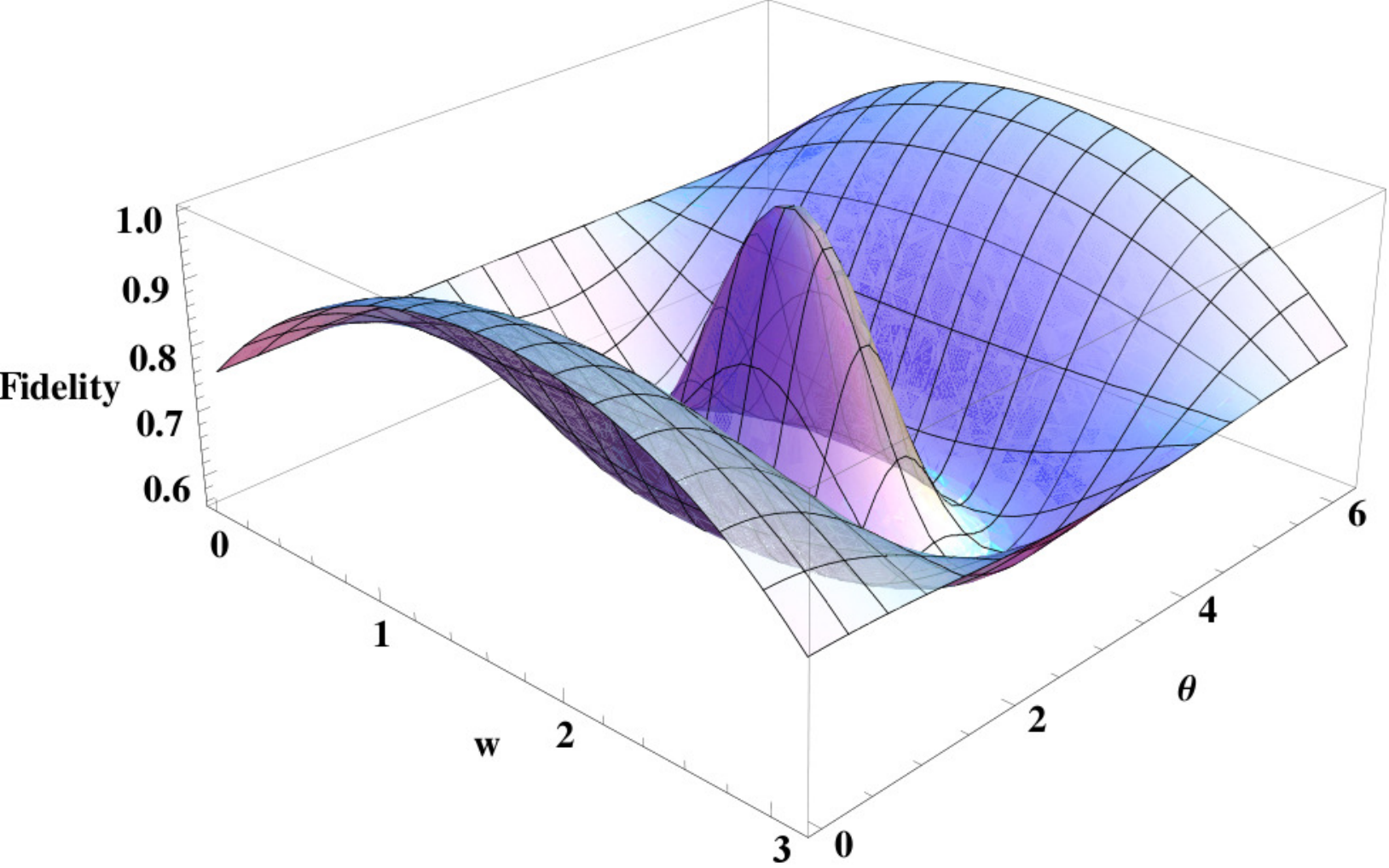} \\
(a) $|\alpha| = 0.6$, $\eta=0.66$ \\
\quad \\
\includegraphics[scale=0.51]{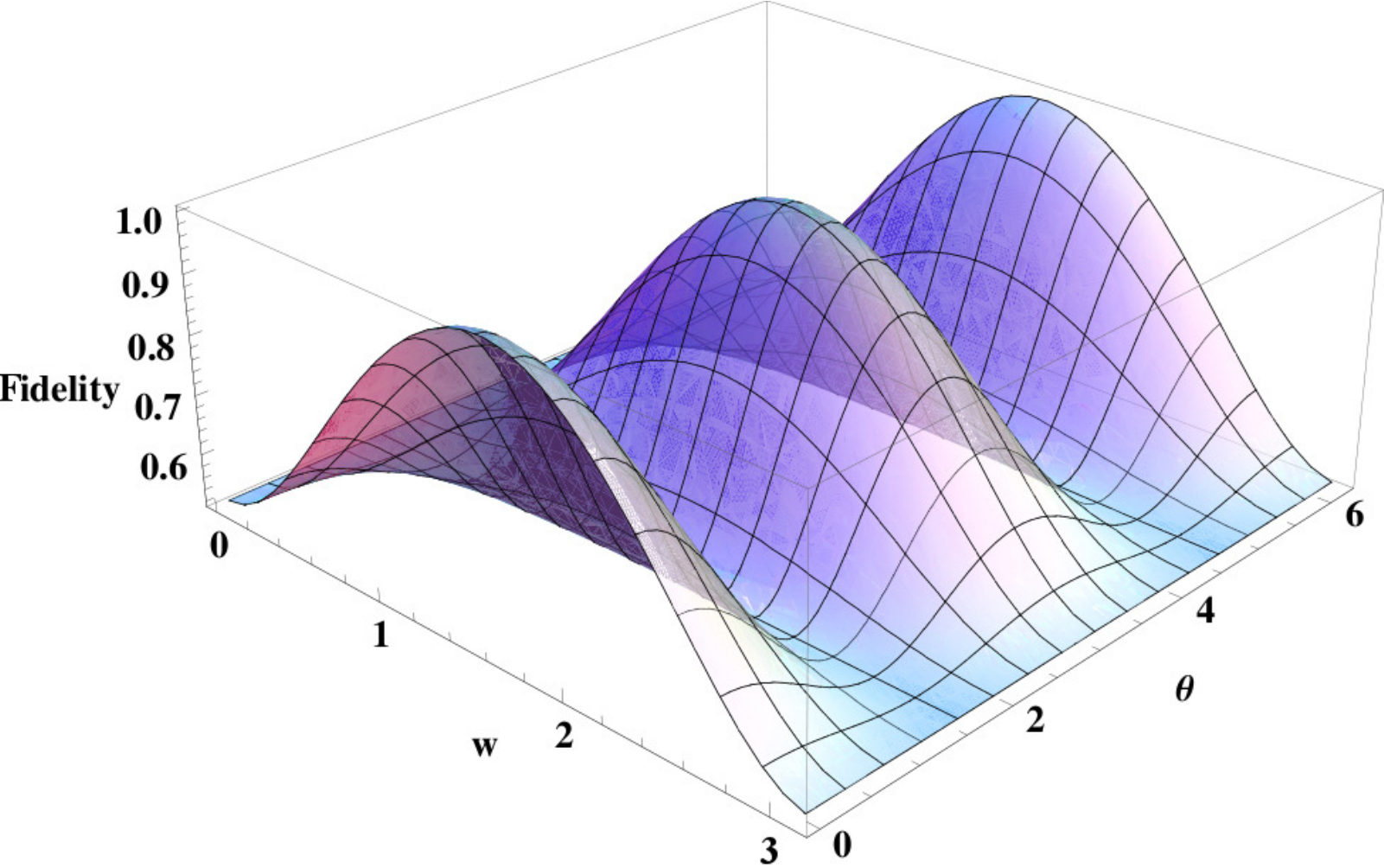} \\
(a) $|\alpha| = 2.4$, $\eta=0.90$
\figcaption{(Color online) Fidelities for different superposition sizes and channel transmissivities: (a) $|\alpha| = 0.6$, $\eta=0.66$ (top), and (b) $|\alpha| = 2.4$, $\eta=0.90$ (bottom right).}
\label{fig:worstcase}
\end{minipage}

\subsection{Codeword Overlap}

The average fidelity between outputs originating from pairs of orthogonal codewords \cite{BenchmarkPaper} is a measure loosely based on the criteria for the existence of error correction codes \cite{KLCriteria} taken in a quantitative sense \cite{DebbieApprox,*BenyApprox}. In an ideal scenario where no errors occur (or assuming perfect correction of all errors is possible), the overlap between the outputs resulting from two orthogonal input states will be zero. In all other cases, the amount by which the distinguishability between two states diminishes establishes a figure of merit with which to evaluate quantum codes. The quantity is defined as \begin{align}
\label{fidoverlap}
F^{CW} = \int dQ \operatorname{Tr} \left[\sqrt{\sqrt{\rho_Q} \rho_{\tilde{Q}} \sqrt{\rho_Q}}\right] \quad , 
\end{align}
where $\rho_Q$ and $\rho_{\tilde{Q}}$ correspond to the outputs originating from a pair of orthogonal input states $|Q\rangle$ and $|\tilde{Q}\rangle$, with $|Q\rangle \perp |\tilde{Q}\rangle$. When considering a non-orthogonal input alphabet, states diametrically opposed in the Bloch sphere should be used instead.

\subsection{Entanglement}

Finally, as in \cite{OurPaper}, recasting a QEC code as an EP protocol,  one can use an entanglement measure as a figure of merit to provide quantitative benchmarks for this code. For the purposes of our analysis, we start with a two-mode maximally entangled state (MES), for instance,
\begin{align}
\label{maxentstate}
|\Phi^-\rangle = \frac{1}{\sqrt{2-2e^{-4|\alpha|^2}}} (|\alpha,\alpha\rangle - |-\alpha,-\alpha\rangle) \quad .
\end{align}
The first mode is kept while the second is sent through the error-correcting scheme, resulting in a state $\rho$ from which one calculates the concurrence \cite{Wooters},
\begin{align}
\label{conceq}
C = \max\{0,\sqrt{\lambda_1}-\sqrt{\lambda_2}-\sqrt{\lambda_3}-\sqrt{\lambda_4} \} \quad.
\end{align}
Here, $\lambda_i$ are the eigenvalues, in decreasing order, of $\rho \tilde{\rho}$, where $\tilde{\rho} = ( \sigma_{y,1} \otimes \sigma_{y,2} ) \rho^{*} ( \sigma_{y,1} \otimes \sigma_{y,2} )$, and $\sigma_{y,i}$ is the Pauli $Y$ operator in the $i$-th mode. 

Alternatively, if $\rho$ is a so-called X state, the concurrence can be found by a simple expression involving the diagonal and off-diagonal elements of the corresponding density matrix \cite{eberly,OurPaper}:
The concurrence of a state described by an X matrix can easily be found by \cite{eberly}
\begin{align}
\label{conX}
	C = 2 \mbox{ max} \left[0,|z|-\sqrt{ad},|f|-\sqrt{bc}\right] \quad ,
\end{align}
where $a$, $b$, $c$, $d$, $f$ and $z$ are the elements of the density operator
\begin{align}
\left(1\otimes\$\right) |\Phi^- \rangle \langle \Phi^-| = 
\left(
\begin{array}{cccc}
a & 0 & 0 & f \\
0 & b & z & 0 \\
0 & z* & c & 0 \\
f* & 0 & 0 & d \\
\end{array} \right) \; .
\label{matx}
\end{align}

\section{Results}

With the above considerations in mind, three regimes will be studied: entirely post-selected; with post-selected (off-line) encoding and deterministic decoding; and with fully deterministic gates. The first regime operates in \textit{Error Detection} mode, requiring two-way communication to address those runs in which one or more of the gates have failed. The other two correspond to operation in true \textit{Error Correction} mode.

\subsection{Fully post-selected gates}

The post-selected regime corresponds to the analysis performed previously \cite{OurPaper}, but is now expanded to include the probabilities of successfully obtaining the desired heralding outcome from the idealized Hadamard gates.

Comparing Figs. 2 and 3, which correspond to approx. 1.8dB and 0.45dB loss regimes, one observes the effect induced by higher transmissivity coefficients, and the increasing benefit of higher-order codes. However, one should take into account the overall probability of success, noting that it diminishes for codes with higher $N$. This is due to the fact that, as the number of modes used in the encoding increases, the size of each superposition arriving at the individual Hadamard gates is reduced accordingly (following $\alpha/\sqrt{N}$). This, in turn, reduces the successful component in the Hadamard transform, as the states become inherently less distinguishable.

\noindent\begin{minipage}{\linewidth} 
\centering%
\includegraphics[scale=0.51]{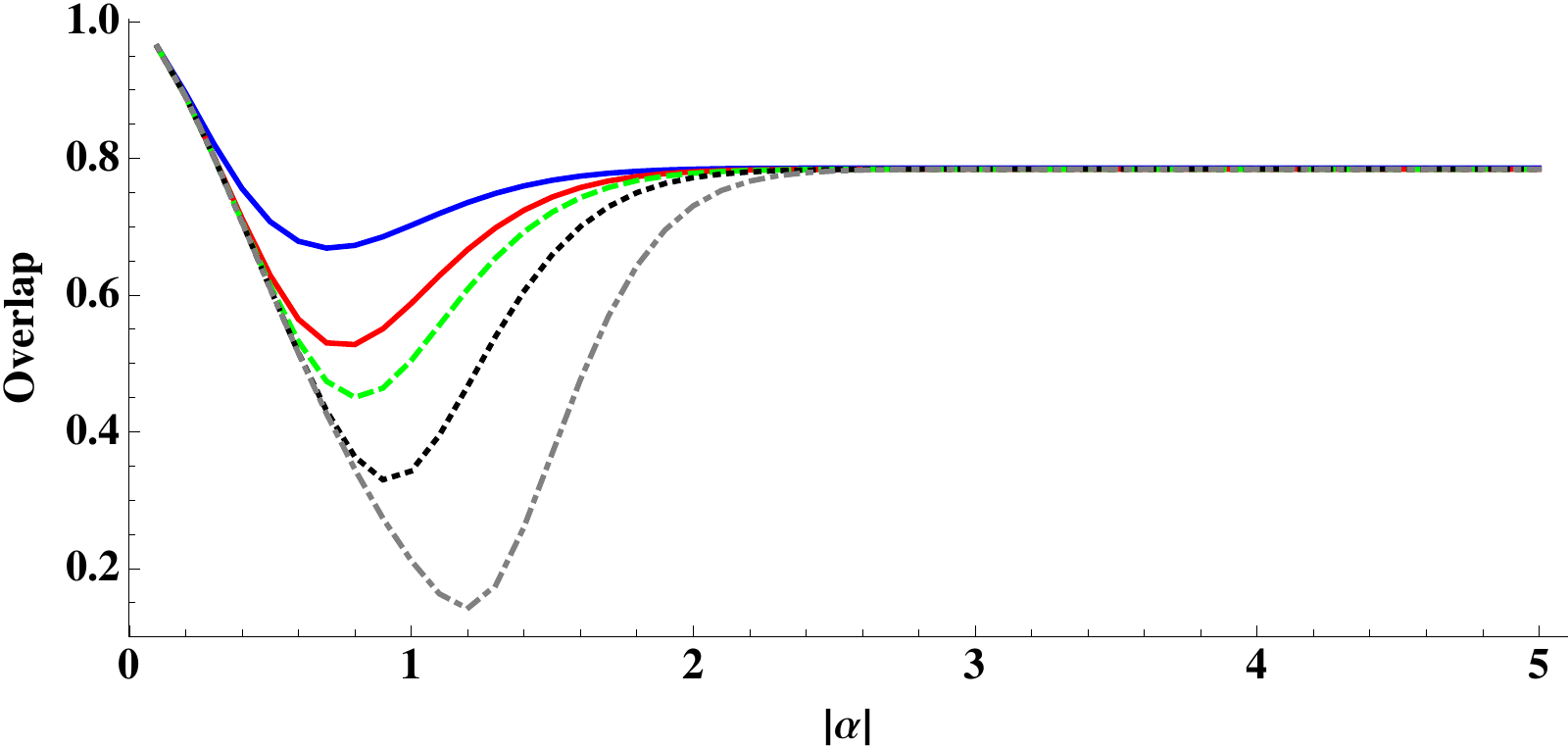} \\
\quad \\
\includegraphics[scale=0.51]{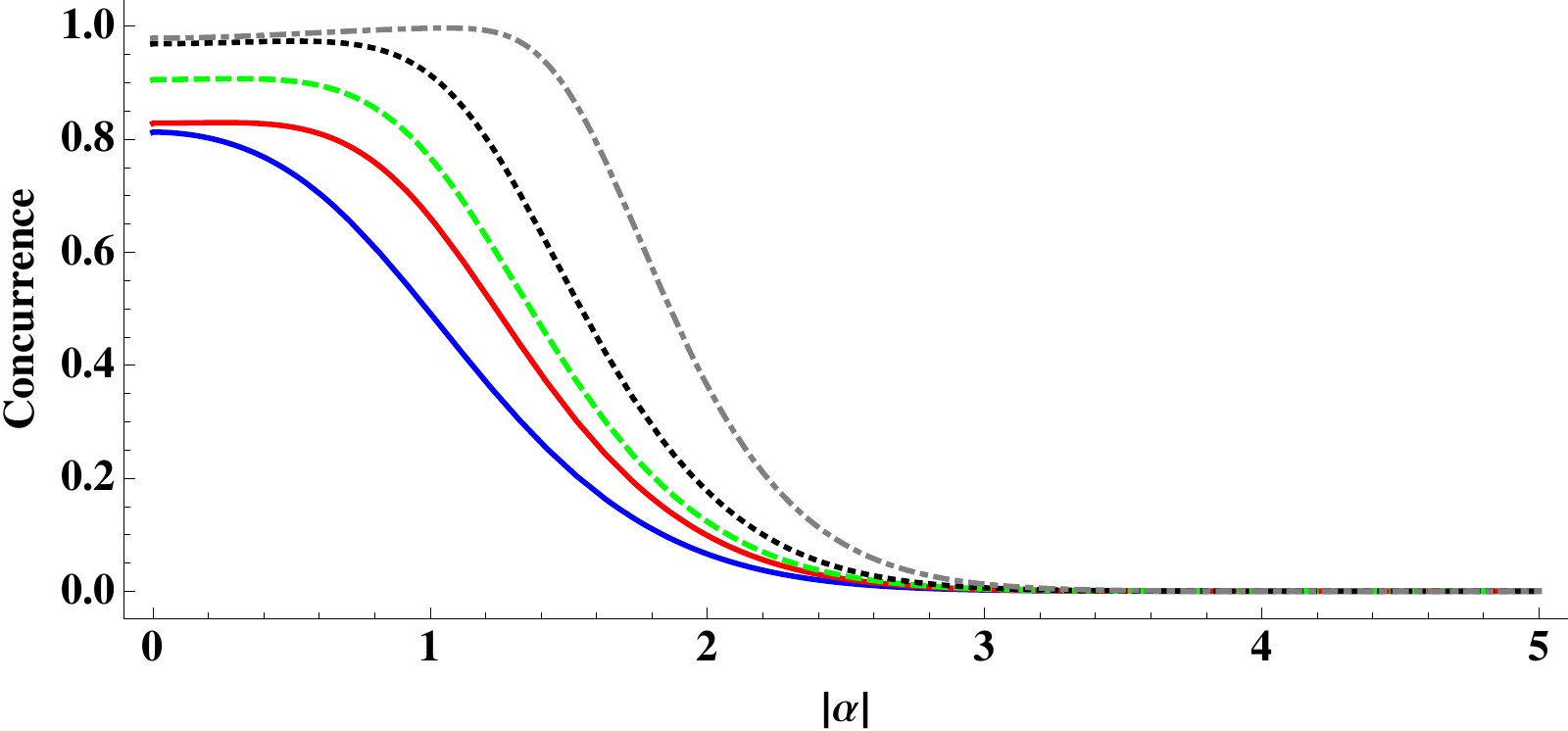} \\
\quad \\
\includegraphics[scale=0.51]{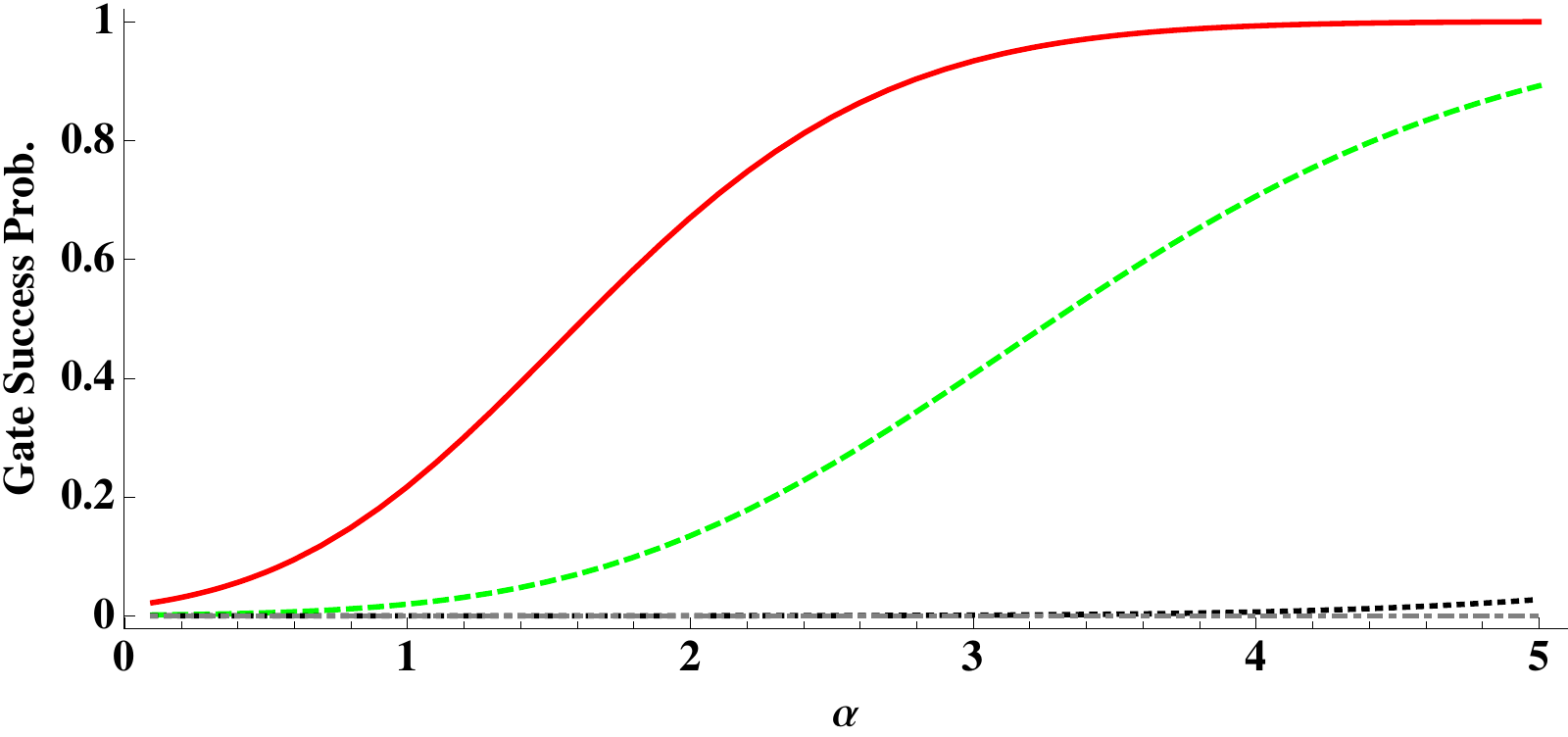} 
\figcaption{(Color online) Codeword overlap (top), concurrence (middle) and gate probability of success (bottom) for the post-selected scheme after transmission through a $\eta=0.66$ channel, as a function of the coherent state size $|\alpha|$. In this and subsequent figures, the following cases are represented: direct transmission (blue, thick line), encoding with 3 (red line), 5 (green, dashed), 11 (black, dotted), and 51 (grey, dashed) qubits.}
\label{fig:postsel066}
\end{minipage}

\begin{minipage}{\linewidth} 
\centering%
\includegraphics[scale=0.51]{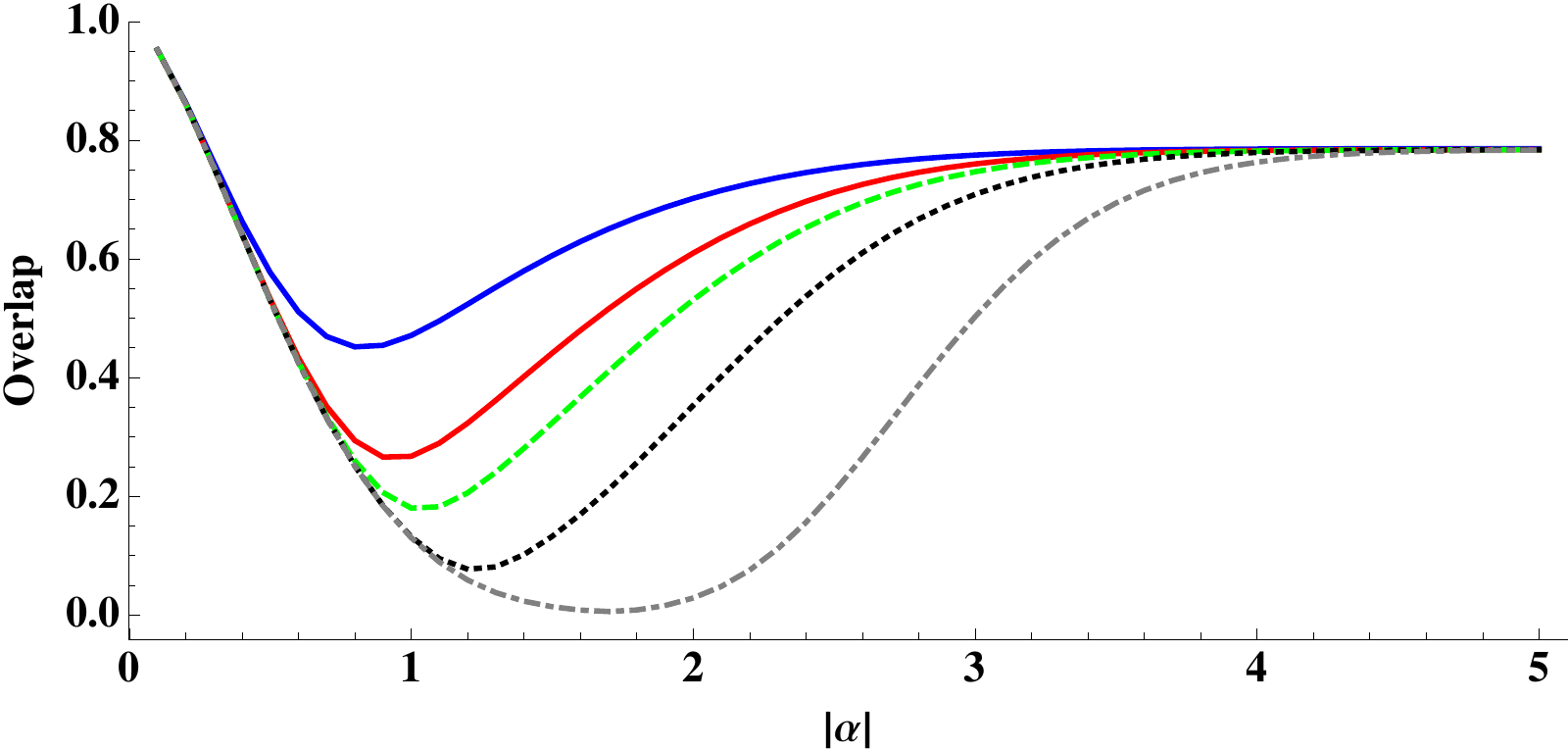} \\
\quad \\
\includegraphics[scale=0.51]{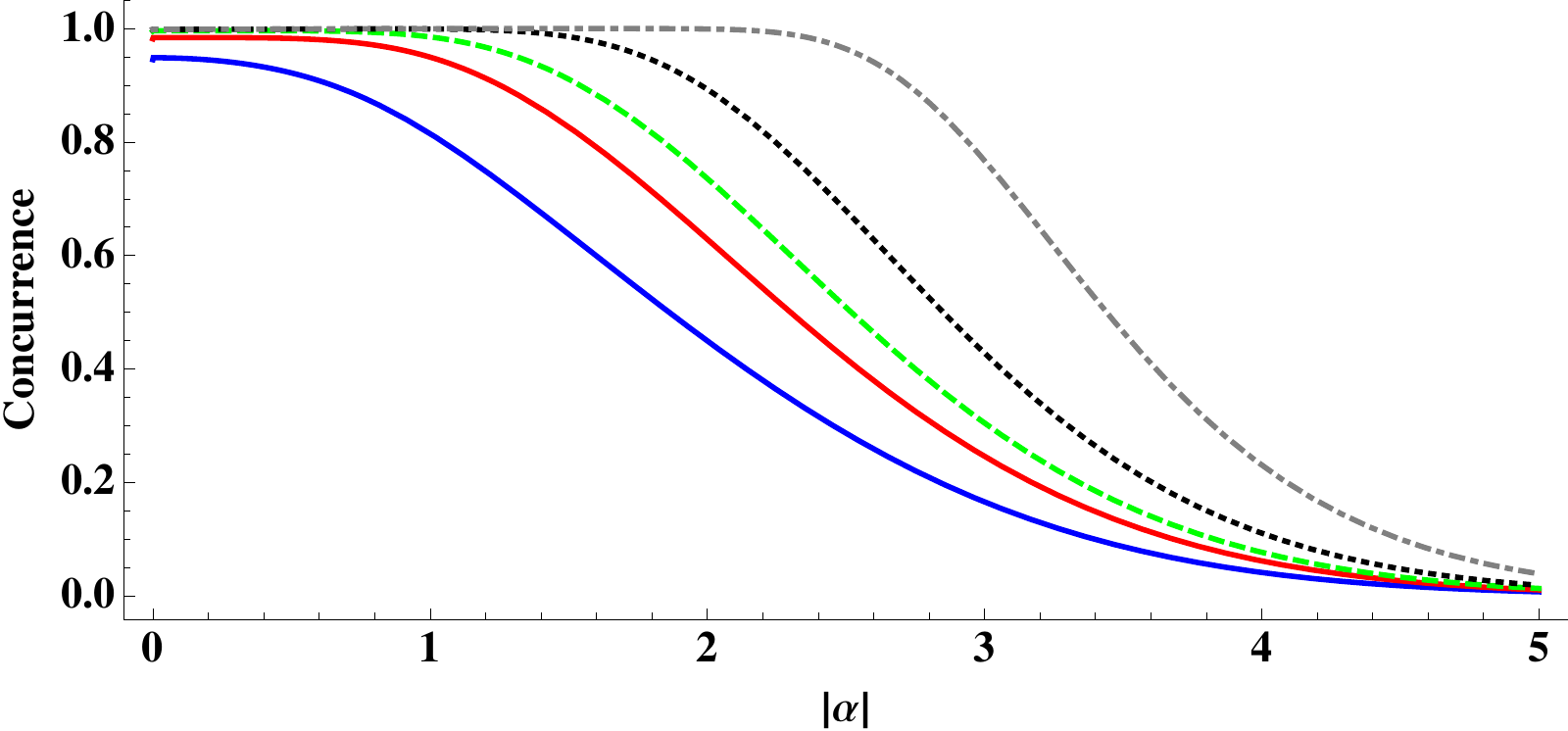} \\
\quad \\
\includegraphics[scale=0.51]{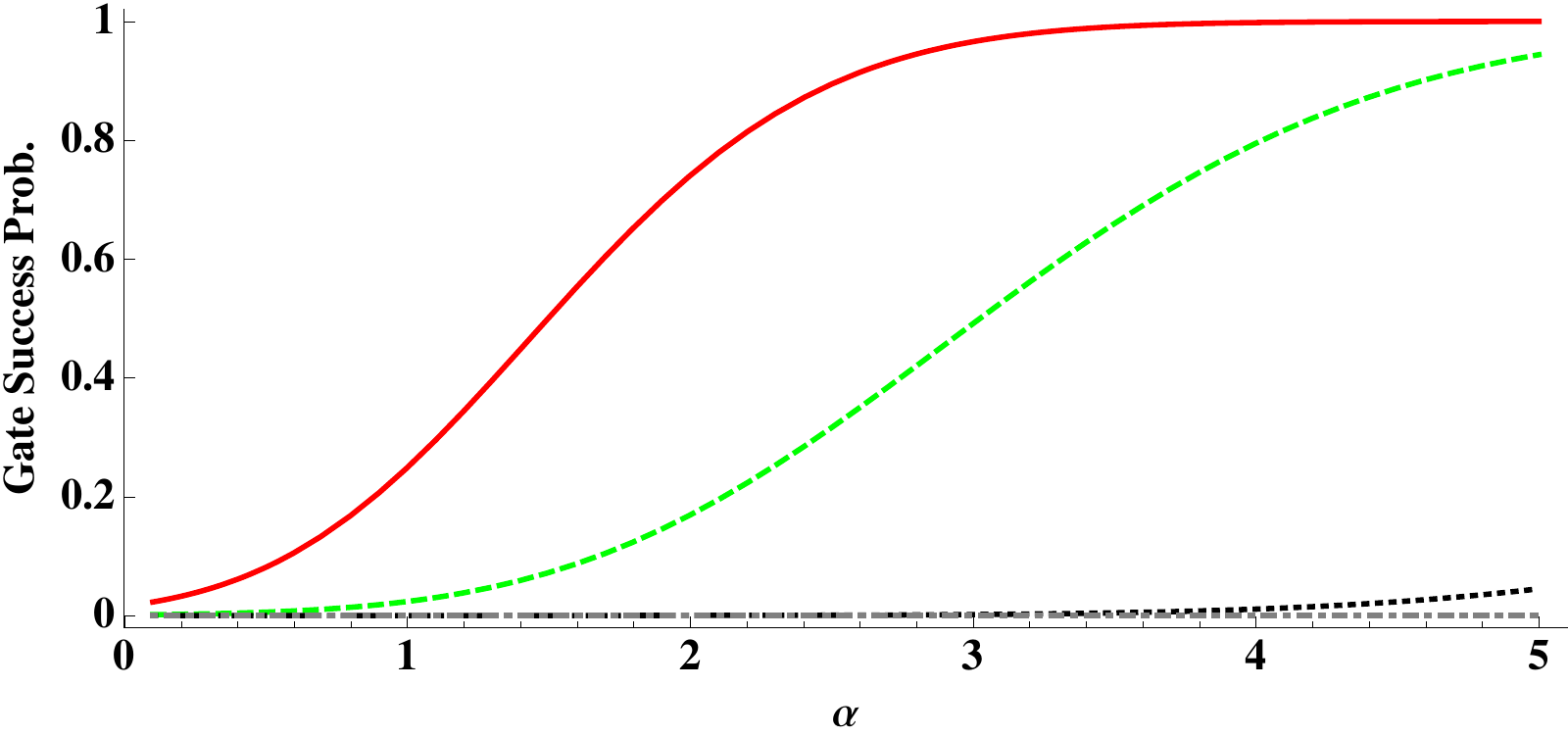} 
\figcaption{(Color online) Codeword overlap (top), concurrence (middle) and gate probability of success (bottom) for the post-selected scheme after transmission through a $\eta=0.90$ channel, as a function of the coherent state size $|\alpha|$. }
\label{fig:postsel090}
\end{minipage}

\subsection{Post-selected encoding}
In this scenario, the encoding is assumed to be available as an off-line resource, with the input qubit only teleported into the encoding when the latter returns a successful heralding signal. 

The competing effect of different repetitions outlined above becomes more explicitly manifest, as the adverse behavior of the gates is now incorporated directly into the three figures of merit considered. In both Figs. 4 and 5 (corresponding, respectively, to $\eta=0.66$ and $\eta=0.90$), the expected \textit{status quo}, that a higher number of repetitions provides quantitatively ``better" figures of merit, seems still to be preserved. Nevertheless, one can already note a dissemblance in that, for a certain range of $|\alpha|$, direct transmission yields more favorable results, hinting at a change of ordering that will become evident in the scenario that follows.

\noindent\begin{minipage}{\linewidth}
\centering
\includegraphics[scale=0.51]{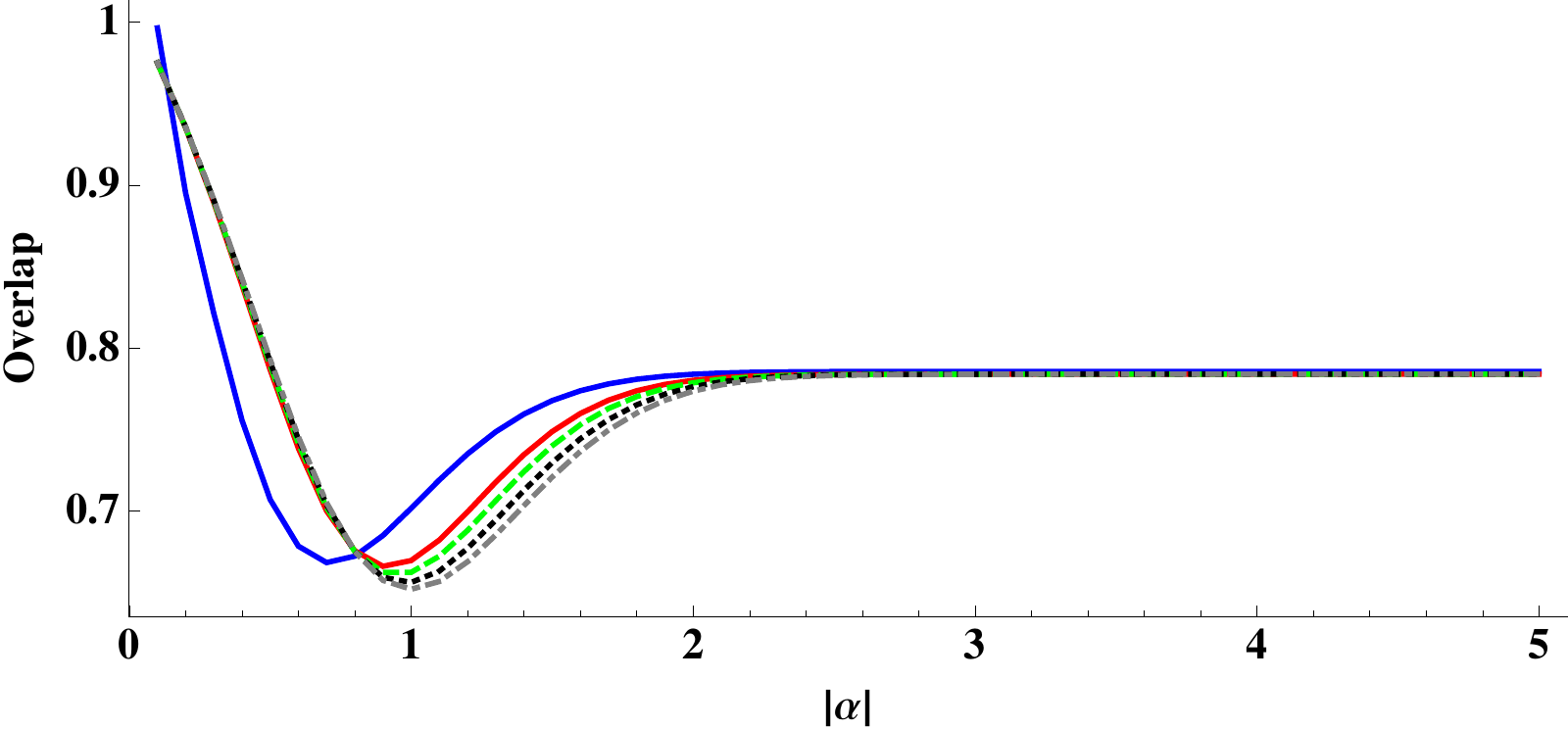} \\
\quad \\
\includegraphics[scale=0.51]{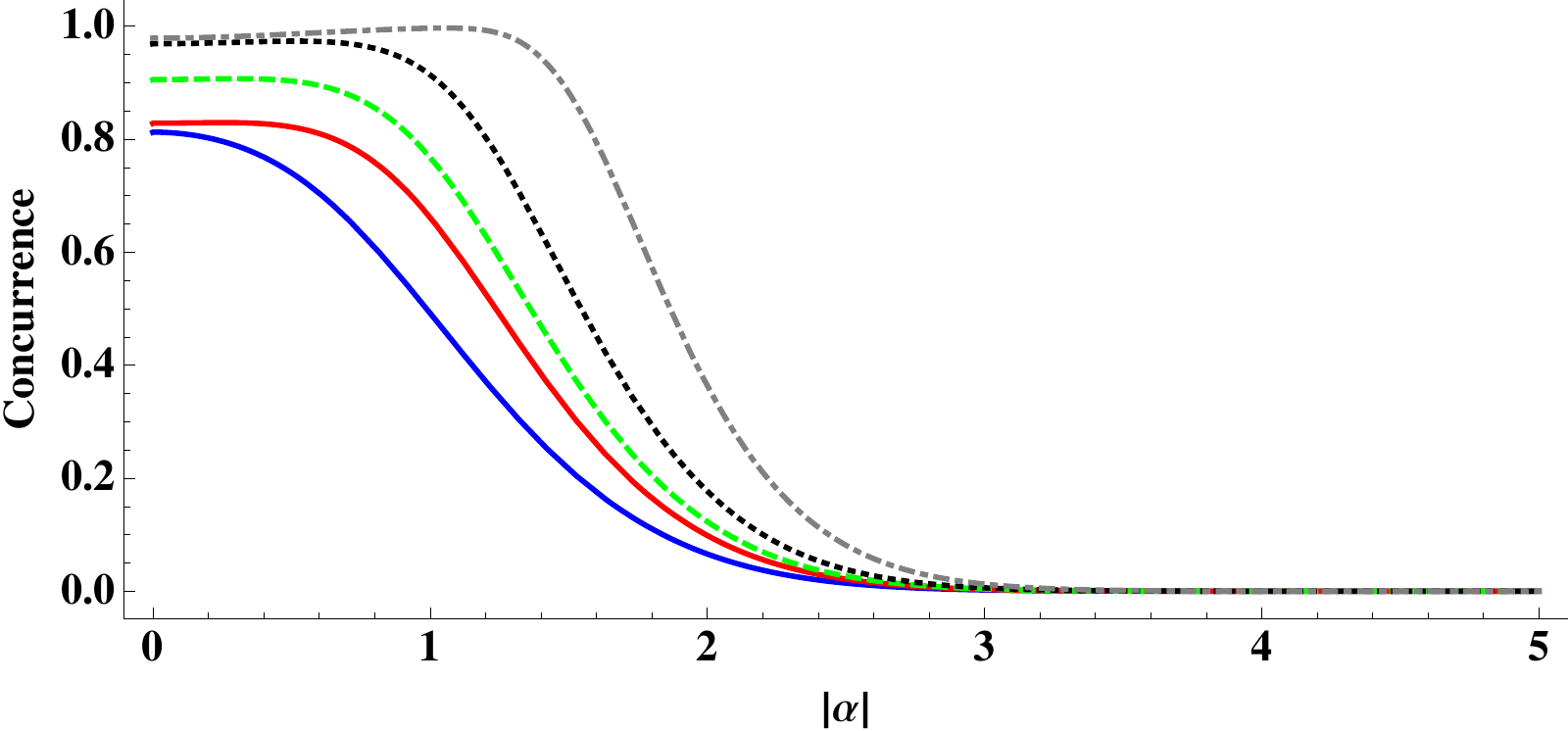}
\figcaption{(Color online) Codeword overlap (top) and concurrence (bottom) for the off-line encoding scheme after transmission through a $\eta=0.66$ channel, as a function of the coherent state size $|\alpha|$.}
\label{fig:semi066}
\end{minipage}

\quad \\

\noindent\begin{minipage}{\linewidth}
\centering
\includegraphics[scale=0.51]{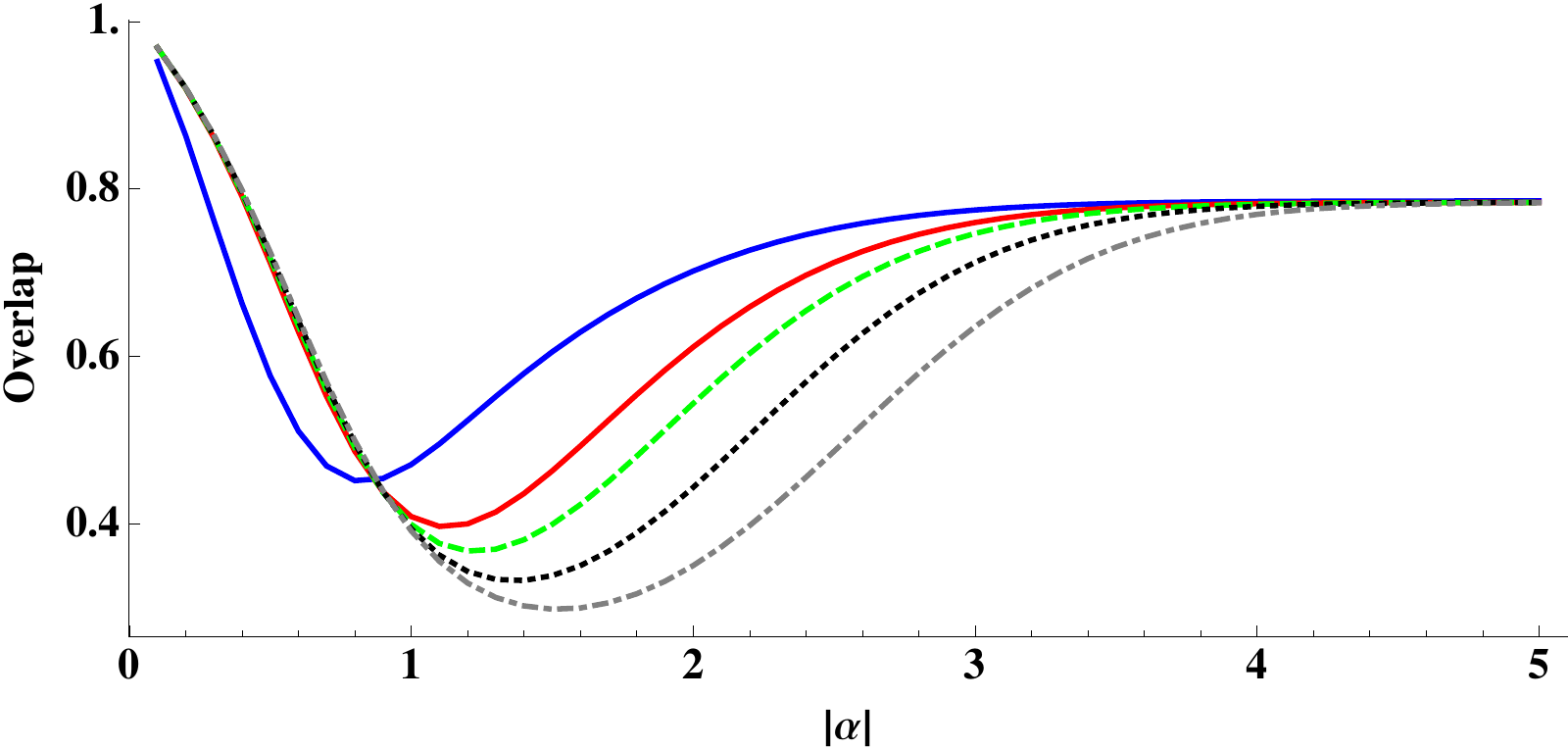} \\
\quad \\
\includegraphics[scale=0.51]{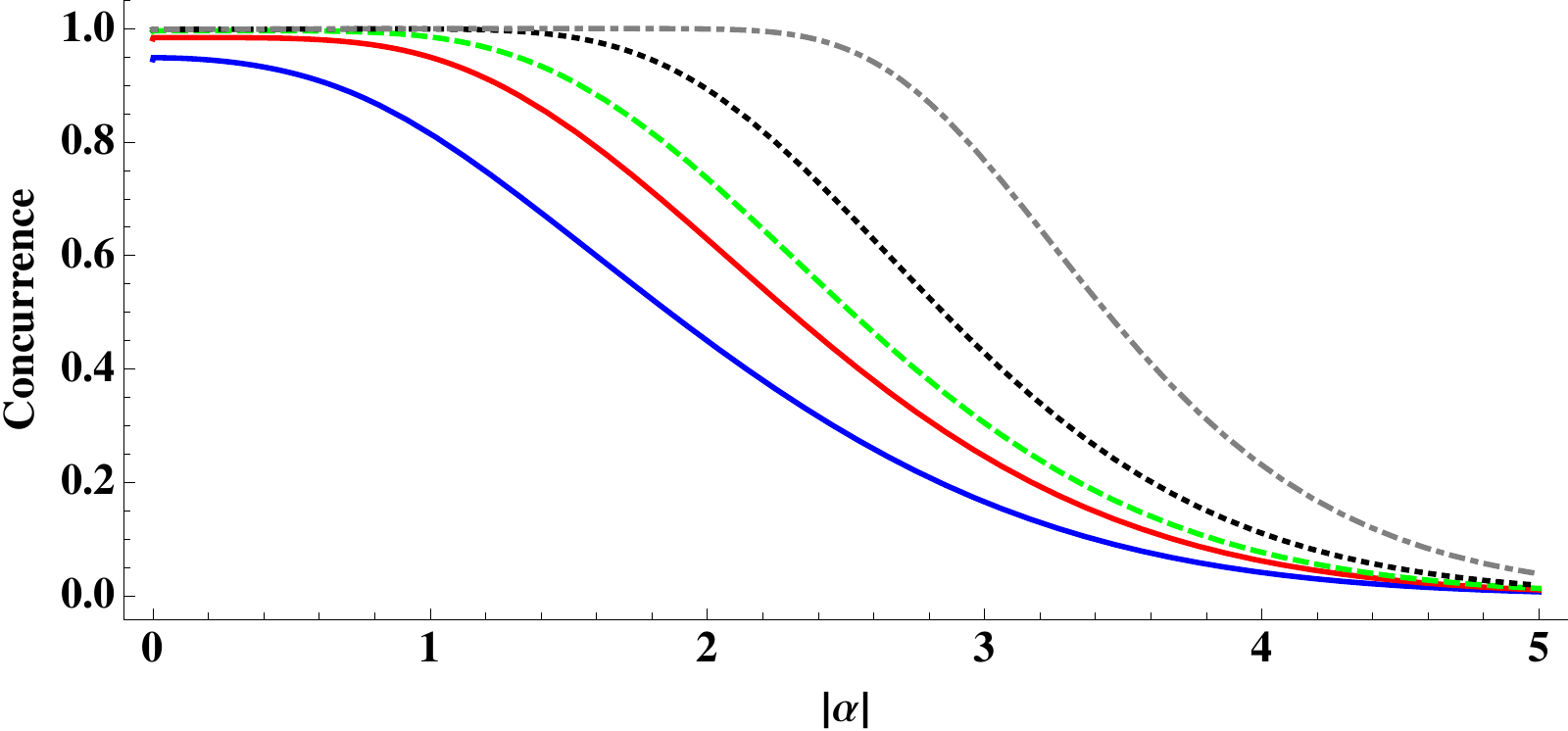}
\figcaption{(Color online) Codeword overlap (top) and concurrence (bottom) for the off-line encoding scheme after transmission through a $\eta=0.90$ channel, as a function of the coherent state size $|\alpha|$.}
\label{fig:semi090}
\end{minipage}

\subsection{Fully deterministic scheme}
Here, deterministic encoding and decoding operations will be employed, relinquishing the usage of post-selected resources. The adverse effect incurred by increasing the number of repetitions is particularly highlighted in regions - determined by the pair ($\alpha$,$\eta$) - where the increased protection granted through further repetitions is suppressed by the growing erroneous component induced by the encoding and decoding operations.

\noindent\begin{minipage}{\linewidth}
\centering
\includegraphics[scale=0.51]{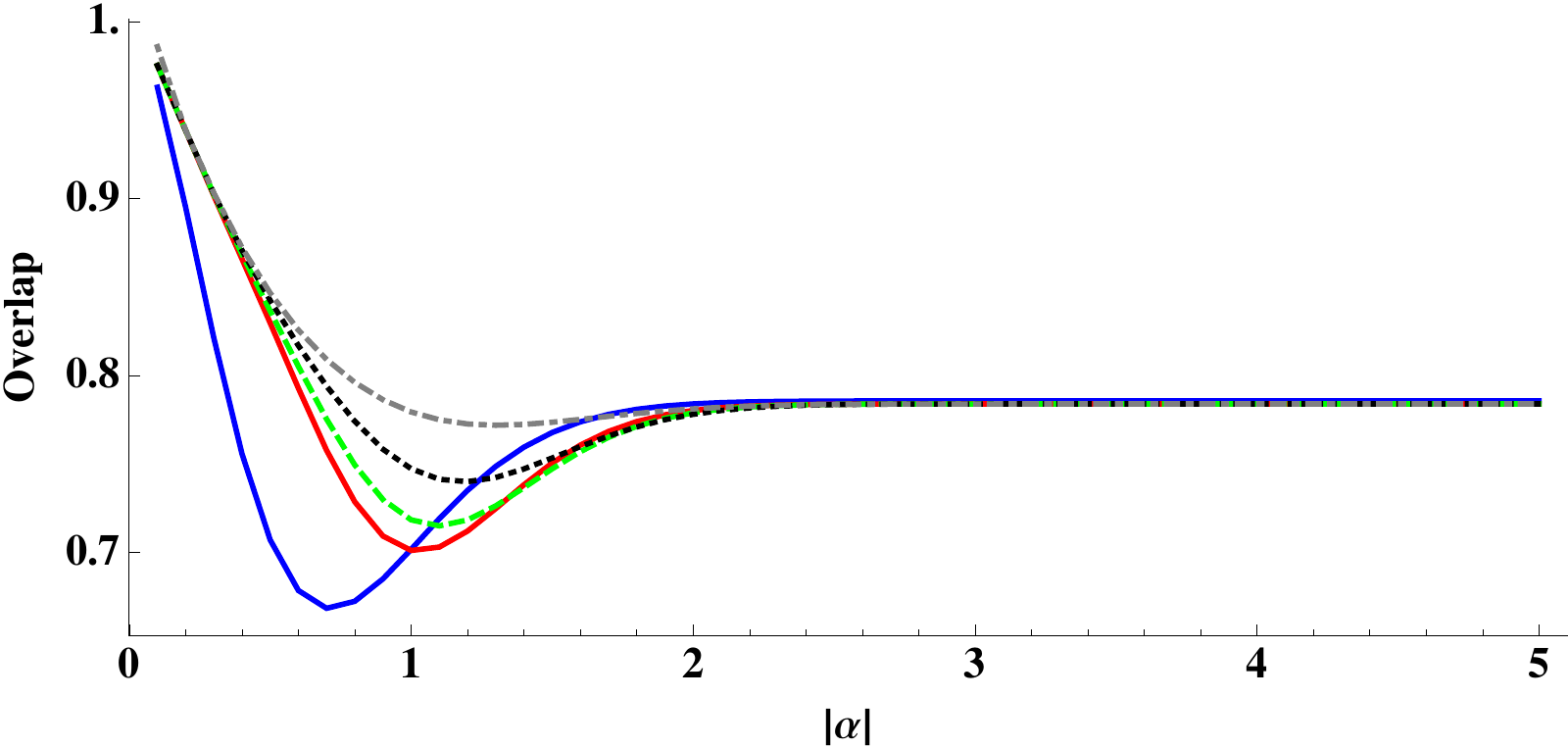} \\
\quad \\
\includegraphics[scale=0.51]{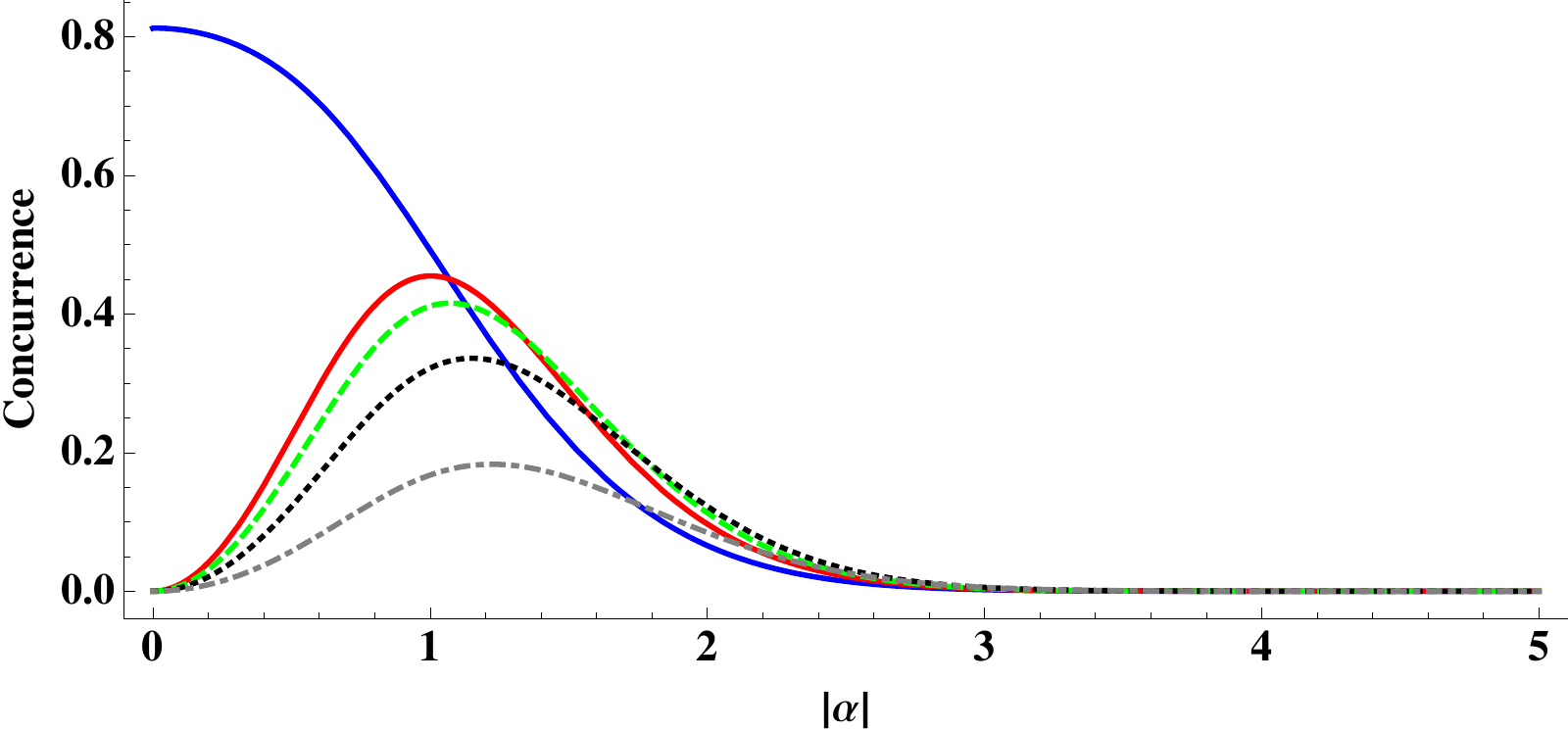}
\figcaption{(Color online) Codeword overlap (top) and concurrence (bottom) for the fully deterministic scheme after transmission through a $\eta=0.66$ channel, as a function of the coherent state size $|\alpha|$.}
\label{fig:full090}
\end{minipage}

\noindent\begin{minipage}{\linewidth}
\centering
\includegraphics[scale=0.51]{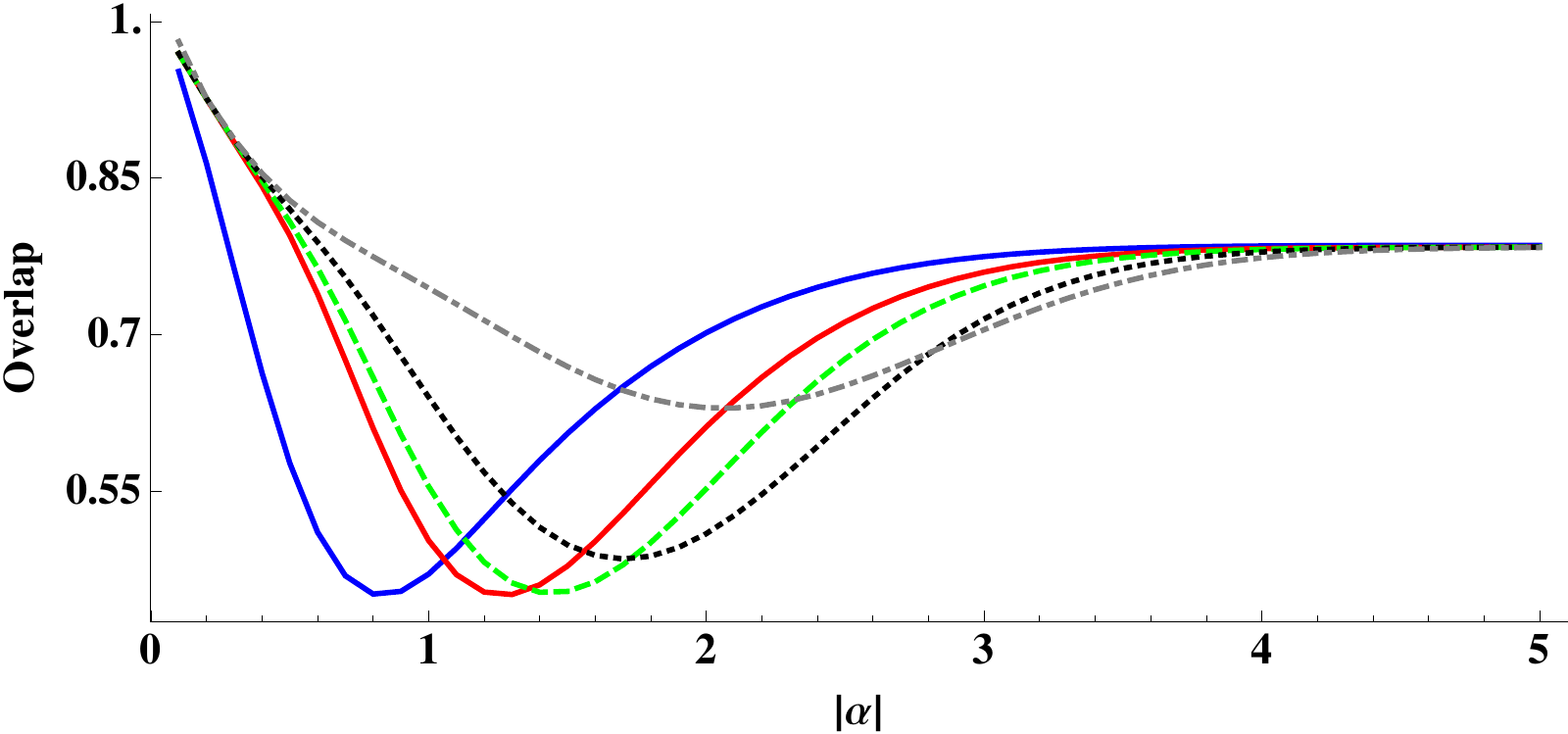} \\
\quad \\
\includegraphics[scale=0.51]{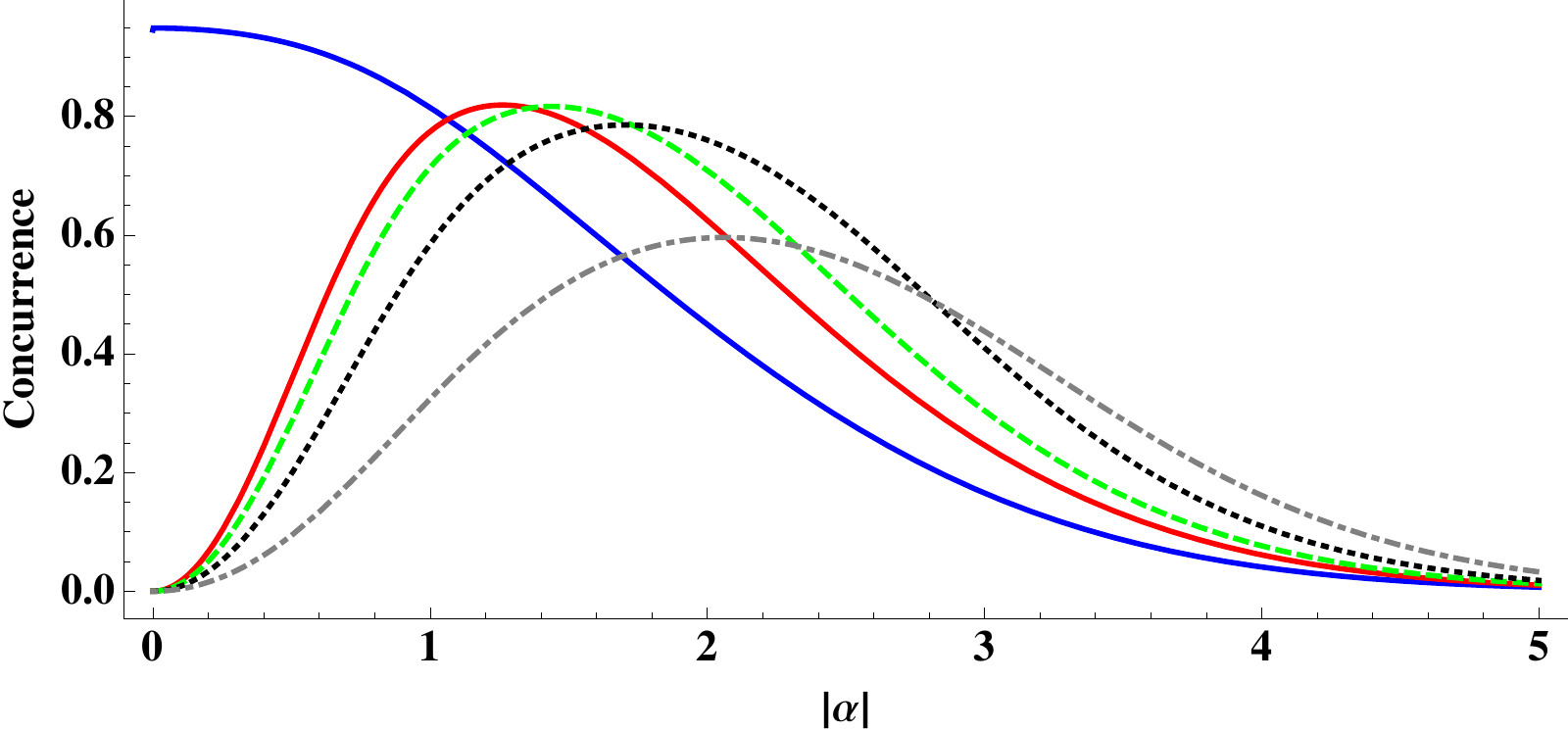}
\figcaption{(Color online) Codeword overlap (left) and concurrence (right) for the fully deterministic scheme after transmission through a $\eta=0.90$ channel, as a function of the coherent state size $|\alpha|$.}
\label{fig:full066}
\end{minipage}

%
%
\pagebreak

In other words, different ($\alpha$,$\eta$) regimes have different optimal $N$; this is evidently seen as the superposition size increases in Fig. 6 and even more so in Fig. 7: Initially, for $\alpha$ up to $\sim 1$, direct transmission is the better alternative. For $\alpha$ between $\sim 1$ and $\sim 1.3$, the 3-qubit code is preferable. In the regime between $\sim 1.3$ and $\sim 1.7$, the five-qubit repetition code produces the best figures of merit. From $\sim 1.7$ to $\sim 2.8$, the 11-repetition code is superior to the alternatives depicted, and from $\sim 2.8$ onwards, the increased protection offered by the 51-qubit code makes this the protocol of choice. Furthermore, also the relative ordering (\textit{i.e.}, the 2nd best ranked, 3rd best, etc) changes.

\section{Discussion and Conclusions}
We have performed a quantitative analysis of a known quantum error correction code, designed to protect quantum bits encoded into coherent-state superpositions. This analysis was performed through distinct quantities: (worst-case) fidelity and codeword overlap, and entanglement, as measured by the concurrence.

A trade-off was observed between the probability of a channel-induced phase-flip and the successful component of the target state after being subject to a non-Gaussian Hadamard, originating from the (non-)orthogonality of the logical alphabet. Since the number of repetitions influences the ``size", and thus the amount of orthogonality, of the states entering the Hadamard gates, a competing behavior was found which hints at an optimal regime where, perhaps somewhat surprisingly, higher-order encodings are not always beneficial. Furthermore, in all regimes considered, when comparing figures of merit across different values of $\eta$ for the same $|\alpha|$ size, the high sensitivity to channel losses is	 explicitly manifest.

By exploring the scheme under different operating modes, the interplay between non-orthogonality and non-unitarity was investigated and a connection with deterministic/error-correction and probabilistic/error-detection regimes was pointed out. This distinction is particularly relevant in the context of high-performance quantum information sharing: Error correction-based quantum  repeaters typically rely on one-way communication to obtain fast transmission rates \cite{QREncoding}.

Our results suggest this class of codes is constrained not only by the experimental difficulty to achieve reasonable coherent-state superposition sizes, but also by the tight tolerances for loss placed on the optical fibers, which significantly limit transmission lengths. Choosing larger amplitudes - proportional to $\alpha \sqrt{N}$ - for the initial CSS qubits, in order to compensate for the increasing failure effect of the non-unitary Hadamard gates with growing encoding levels, adds extra expensive, or even unfeasible, resources to the scheme. Nevertheless, such codes may still find application outside of the communication context, such as, for instance, in the short-scale lengths inside an optical quantum computer based upon CSS-qubit encoding.

\subsection*{Acknowledgments}
The authors thank Carlo Cafaro for useful discussions. Support from the Deutsche Forschungsgemeinschaft (DFG) through its Emmy Noether Program, and from the Bundesministerium f\"ur Bildung und Forschung (BMBF) by means of the HIPERCOM project, is gratefully acknowledged.

\bibliography{bibio-ep2}

\begin{thebibliography}{25}
\bibitem [{Notei()}]{Notei}%
  \BibitemOpen
  \bibinfo {note} {The QND measurement is useful for two reasons: first, it
  forces a flip/no flip behavior, which enables one to combine the Hadamard
  failure probability with the phase flip probability $p_e$ induced by
  amplitude damping channel, resulting in a global success probability for the
  code based on the triplet ($\alpha $,$\eta $,$N$). Second, by removing the
  off-diagonal elements in the resulting density matrix, the entanglement can
  be calculated by more effective means, see section VI. C.}\BibitemShut
  {Stop}%
\end{thebibliography}

\end{document}